\documentclass{mpe_report}

\usepackage{psfig,graphicx,epsfig}
\usepackage{color}
\usepackage{amsmath,amssymb,epic,eepic,array}

\begin{document}

\pagenumbering{arabic}
\setcounter{page}{16}

 \renewcommand{\FirstPageOfPaper }{ 16}\renewcommand{\LastPageOfPaper }{ 19}\newcommand{\araa}{\em Ann. Rev. Astron. Astrophys.}
\newcommand{\aap}{\em Astr. Astrophys.}
\newcommand{\aaps}{\em Astr. Astrophys. Supp.}
\newcommand{\nat}{\em Nature}
\newcommand{\aj}{\em Astronom. J.}
\newcommand{\apj}{\em Astrophys. J.}
\newcommand{\apjl}{\em Astrophys. J. Lett.}
\newcommand{\apjss}{\em Astrophys. J. Supp.}
\newcommand{\mnras}{\em Mon. Not. R. Astr. Soc.}

\title{Correlations between Pulsed X-ray Flux and Radio Arrival Time in 
the Vela Pulsar}
\author{A. Lommen\inst{1}, J. Donovan\inst{1,2},  
C. Gwinn\inst{3},
Z. Arzoumanian\inst{4,5}, A. Harding\inst{4}, 
M. Strickman\inst{6}, R. Dodson\inst{7},
P. McCulloch\inst{8}, and D. Moffett\inst{9}}
\institute{Department of Physics and Astronomy, Franklin and Marshall College, Lancaster, Pennsylvania
\and Department of Astronomy, Columbia University, New York, New York
\and Department of Physics, University of California, Santa Barbara, California
\and NASA Goddard Space Flight Center, Greenbelt, Maryland
\and Universities Space Research Association
\and Code 7651.2, Naval Research Laboratory, Washington, DC
\and Observatorio Astron\'omico Nacional, Madrid, Espa\~na
\and University of Tasmania, Tasmania, Australia
\and Furman University, Greenville, South Carolina}

\authorrunning{Lommen et al.}

\maketitle

\begin{abstract}
We report the results of simultaneous observations of the Vela pulsar in 
X-rays and 
radio from the RXTE satellite and the Mount Pleasant Radio Observatory in Tasmania. 
We sought correlations between the Vela's X-ray and radio flux densities and radio
arrival times on a pulse by
 pulse basis. We found significantly higher flux density in
Vela's main X-ray peak
during radio pulses
that arrived early.
This excess flux shifts to the `trough' following the 2nd X-ray peak during radio pulses that arrive later.
We suggest that the mechanism producing the radio pulses is intimately connected to the mechanism producing X-rays.
Current models using
resonant absorption in the outer magnetosphere as a cause of the radio emission, and less 
directly of the X-ray emission, are explored as a possible explanation for the correlation.
\end{abstract}

\section{Introduction}

The complexity of Vela's spectrum allows for the possibility of
both polar cap \cite{Daugherty96} and outer-gap \cite{Cheng00} models of emission,  
but only a subset of the models suggest a connection between the 
radio and X-ray emission. 
For example \cite{Harding02} suggest that scattering of soft X-ray photons
in a polar cap model could explain the observed optical spectrum of Vela
and also provide for the alignment of the radio peak with a very soft
X-ray peak, which they observe.
The connection 
between X-ray and radio emission that we
offer in this article favors models such as this, which connect 
far reaching ends of the electromagnetic spectrum.

Additional experiments linking pulsar emission in the radio and X-ray regimes
have been performed by Cusumano et al. (2003)\nocite{Cusumano03} and 
Vivekanand (2001)\nocite{Vivekanand01}. Cusumano et al. show that in PSR 
B1937+21 there is close phase alignment between X-ray pulses and giant radio
pulses, suggesting a correlation in their emission regions. The work by 
Vivekanand, on the other hand, explains the connection
between the Crab pulsar's radio and X-ray flux variations as a result
of temporal variations in the number of ``basic emitters" or in 
the basic emitters' degree of coherence.

Giant radio pulses have not been shown to exist in Vela, but
Johnston et al. (2001, hereafter J01)
\nocite{Johnston01} discovered `giant micropulses' in the Vela pulsar,
which have flux densities no more than ten times the mean flux density
and have a typical pulse width of $\sim 400 \mu s$.
These micropulses may explain the results of 
\nocite{Krishnamohan83} Krishnamohan \& Downs (1983, hereafter KD83)
who found that the strongest radio pulses arrive earlier than the averaged profile.

By doing a pulse-by-pulse analysis 
of the Vela pulsar in X-ray and radio wavelengths, we will show in this paper that 
the emission mechanisms creating the Vela pulsar's 
X-ray and 
radio flux densities must be related. 
We will discuss the X-ray and radio observations in 
\S \ref{sec.observations}, our analysis in 
\S \ref{sec.analysis}, the effects of scintillation in 
\S \ref{sec.scintillation}, a discussion of interpretations in 
\S \ref{sec.discussion}, theoretical predictions as they relate to our results in 
\S \ref{sec.theory}, and finally our conclusions and related future work in 
\S \ref{sec.conclusion}.

\section{Observations}
\label{sec.observations}
Our data consist of 74 hours
of simultaneous radio
and X-ray observations taken over three months at  
the Mount Pleasant Radio Observatory in Tasmania 
and with the RXTE satellite. 

The radio data were acquired during 12 separate
observations using the 26m antenna at a frequency
of 990.025 MHz between 30 April and 23 August, 1998. 
All individual pulses from Vela are detectable, and the
pulse height, integrated area, and central time of arrival (for the 
solar-system barycenter)
were calculated from cross-correlation with a high signal to noise template in
the usual fashion.

The X-ray data were taken 
during the same three months, yielding 265 ks of
usable simultaneous observation. For the purposes of this project, 
only top-layer data from 
RXTE's Proportional Counter Units (PCUs) in 
Good Xenon mode in the energy range of 2-16 keV were used. Other filtering 
parameters included were standard RXTE criteria: elevation was greater than 10 degrees, offset
was less than 0.02 degrees, the data were taken with at least 3 PCUs on, time since SAA  
was at least 30 minutes, and electron0 was less than 0.105.  

\section{Analysis} 
\label{sec.analysis}
We filtered the X-ray photon arrival times and transformed them
to the Solar System Barycenter (SSB) using the standard FTOOLS
\cite{Blackburn95} package. 
We calculated the pulsar phase at the time of each X-ray
photon, 
and matched each X-ray photon with the radio pulse that arrived at the
SSB at the same time.  
We then compared pulse profiles for X-rays segregated according to
the arrival time of the radio pulse.

We filtered the X-ray photon arrival times and transformed them
to the Solar System Barycenter (SSB) using the standard FTOOLS
\cite{Blackburn95} package. 
We calculated the pulsar phase at the time of each X-ray
photon, using the radio pulsar-timing program TEMPO$^1$,
and the ephemeris downloadable from
Princeton University\footnote{See http://pulsar.princeton.edu/tempo}. 
We matched each X-ray photon with the radio pulse that arrived at the
SSB during the same turn of the pulsar.  The precise period of time 
associated with each
radio pulse was given by our best model for the pulse arrival time,
$\pm 0.5\times$ the instantaneous pulse period calculated via the
model.  Photons arriving on the borderline were associated with the
earlier pulse.
We then compared pulse profiles for X-rays segregated according to
the corresponding radio pulse arrival time.

\begin{figure}[htbp!]
\centerline{\psfig{file=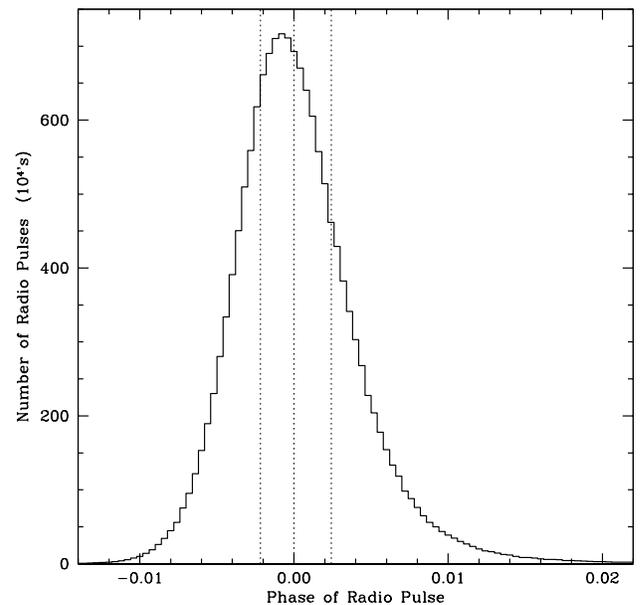,width=8.8cm, clip=}}
\caption{The total number of radio pulses vs phase relative to a predictive
long-term timing model. The 
dotted lines show the boundaries of the four bins that were used to make
the four profiles shown in Figure \ref{fig.hists}.
\label{fig.phasedist}
}
\end{figure}

Radio pulses arrive at a range of times around the predicted arrival time,
as KD83 found.
The histogram of residual arrival times for radio pulses,
relative to the prediction of our best model, 
is shown in
Figure \ref{fig.phasedist}.  
We divided all of the pulses into four quartiles, 
by the residual phase of the radio pulse,
with equal numbers of pulses
in each quartile.
Figure \ref{fig.phasedist} shows our division of the residual phase of the radio pulse into the
quartiles. 
We called the quartiles early, medium-early, medium-late, and late,
according to the arrival time of the radio pulse.

\begin{figure}[htbp!]
\centerline{\psfig{file=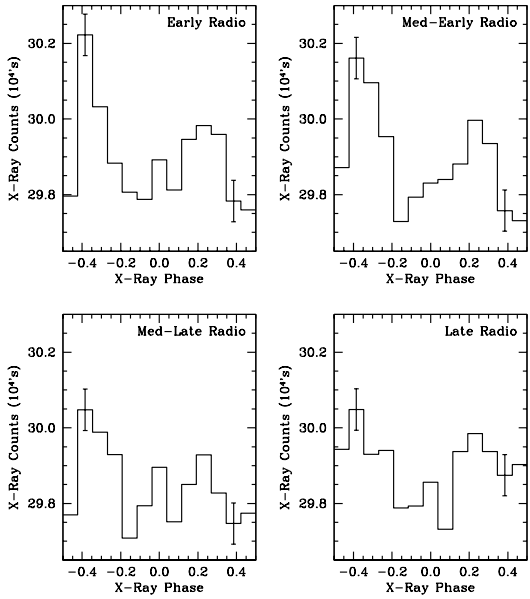,width=8.8cm, clip=}}
\caption{Full-period X-ray profiles for
          photons detected during radio pulse arrival times falling
          in the four quartile bins shown in 
Figure \ref{fig.phasedist}. The radio peak falls at phase 0.5.
\label{fig.hists}
}
\end{figure}

We formed an X-ray pulse profile for each of the quartiles of radio-pulse
arrival times,
from the X-ray photons associated with each.
Figure \ref{fig.hists} shows the four resulting X-ray profiles with
errors determined by counting statistics.
The X-ray profiles are significantly different;
the pulse changes in shape among the quartiles.
In particular it appears that the first X-ray pulse is sharper
and stronger during ``early" radio pulses.  Also, it appears that the
trough after the 2nd X-ray peak around phase 0.4 
(hereafter called simply `the trough') 
gets ``filled in" in late quartiles.

\begin{figure}[htbp!]
\centerline{\psfig{file=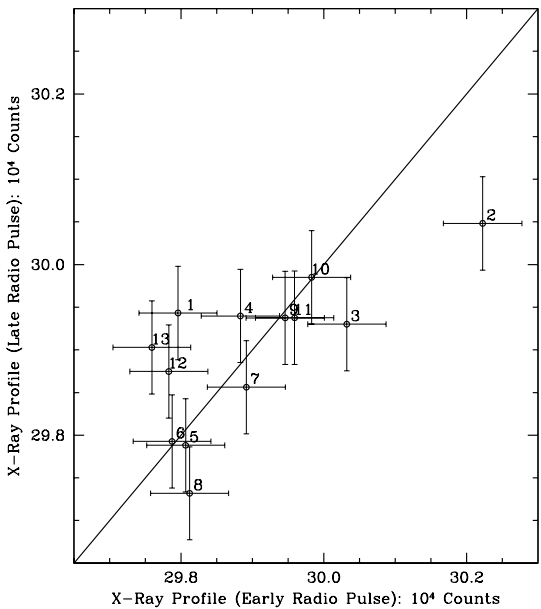,width=8.8cm, clip=}}
\caption{Measured counts in the late radio phase quartile vs counts in the
early radio phase quartile.  X-ray phase bins are numbered sequentially beginning
with 1.
\label{fig.bin0vsbin3}
}
\end{figure}

In order to graphically analyze the changes in pulse shape,
we plot
the profiles in different quartiles against one another in
Figure \ref{fig.bin0vsbin3}.
The solid diagonal line shows where the points would
lie if the profiles were identical.
As the figure shows, Bin 2 (the tallest bin in the 1st X-ray
peak) is significantly stronger in the 
early quartile than in the late quartile.
In addition, although less noticeable, Bins 12, 13 and 1 (the trough)
are significantly
weaker in the early quartile than in the late.

\begin{figure}[htbp!]
\centerline{\psfig{file=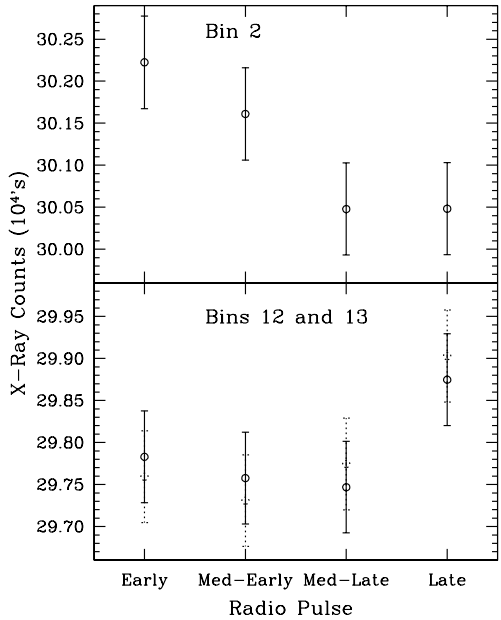,width=8.8cm, clip=}}
\caption{The height of Bin 2, and of the dip in Bins 12 and 13, vs the
lateness of the quartile.
\label{fig.lateness}
}
\end{figure}

Overall, the data suggest that 
sharper X-ray pulses are associated with earlier radio pulses.
In Figure \ref{fig.lateness} we plot the height of the peak in Bin 2, and of the trough in Bins 12 and 13,
against the lateness of the quartile.
The figure shows a systematic trend of a weaker peak toward late arrivals.
The increase in photon rate in the trough makes up for much of the loss
of X-ray photons in the peak.

\section{Scintillation}
\label{sec.scintillation}

In contrast to effects intrinsic to the pulsar, scintillation
is unlikely to produce the observed association, because it
does not affect X-rays; scintillation might erase such a correlation
but it cannot introduce it. 
Nonetheless, we used a number of techniques to ensure that our
result was not produced by scintillation including subtracting the average
residual from each 5-minute span of data, and defining the four quartiles
based on both 5-minute and 1-hour spans of data.  More details can be
found in \cite{Lommen06vela}.  In summary, none of these techniques changed
our results significantly.

\section{Discussion}
\label{sec.discussion}

J01 showed that the giant micropulse emission occurs about 1 ms before the radio peak, 
so it is realistic to consider the possibility that the giant micropulse emission
is primarily responsible for the early arrival of the radio pulse. The hypothesis
is consistent with the timescale of the ``early-ness":
the full-width at half-maximum of the histogram of arrival time residuals
shown in Figure \ref{fig.phasedist} is about 0.5 ms. 
However, the more detailed analysis performed in \cite{Lommen06vela}  
shows that the effect cannot be explained by the occurrence of giant  
micropulses.

\section{Theoretical Predictions}
\label{sec.theory}
The results described above imply a connection between the radio and X-ray emission mechanisms 
for Vela that is not consistent with outer gap models. In these models, the high energy emission 
results from a gap connection to the pole opposite from that producing the radio emission. It is not 
clear how a correlation could exist between the radio and high energy regimes in these models.
Petrova \nocite{Petrova03} 
(2003, and 
references therein), however, offers a detailed model that could explain the 
correlation. She states that the non-thermal optical
emission of rotation-powered pulsars should be considered as part of the 
broadband high-energy emission.
Her models suggest that resonant absorption of radio emission 
from the outer magnetosphere 
leads to an increase in the pitch angles and momenta of the secondary pairs, 
which then leads to optical and higher energy emission by spontaneous synchrotron radiation.

\section{Conclusions and Future Work}
\label{sec.conclusion}
We have significant evidence linking features of Vela's X-ray emission
with features of its radio emission.  We find that X-ray pulses associated with
early radio pulses show stronger emission at the main X-ray peak which is the sharper
of the two.  Similarly X-ray pulses associated with later radio pulses show stronger
emission at the 
trough following the 2nd X-ray peak.
We conclude that there is a close relationship
between X-ray and radio emission in the Vela pulsar.

We plan to further characterize the relationship between the radio and
high energy emissions of pulsars to identify their origins and
constrain magnetospheric models. In particular, we will explore the
dependence of radio-to-X-ray correlations on the radio frequency and
polarization properties of individual Vela pulses, both of which carry
information about emission altitudes. Similar observations of other
pulsars also promise useful insights as probes of different magnetic
field strengths and emission/viewing geometries.

\begin{acknowledgements}
Many thanks to Michael Kramer, Wim Hermsen, David Helfand and Paul Ray for helpful 
comments and to the Anton Pannekoek Institute at the University of Amsterdam for their 
hospitality, 
particularly Ben Stappers, Simone Migliari, Tiziana Di Salvo and Russell Edwards for 
useful discussions. JD and AL gratefully acknowledge the financial support of Franklin and 
Marshall College and a Research Corporation Grant in support of this work. 
ZA was supported by NASA grant NRA-99-01-LTSA-070. 
CRG acknowledges support of the National Science Foundation.
\end{acknowledgements}

           \clearpage


\begin{thebibliography}{}

\bibitem[(Blackburn 1995)]{Blackburn95}
{Blackburn}, J.~K. 1995, In ASP Conf. Ser. 77: Astronomical Data Analysis 
Software and Systems IV, R.~A. {Shaw}, H.~E. {Payne}, and J.~J.~E. {Hayes}, 
eds., p. 367

\bibitem[({Cheng}, {Ruderman}, \& {Zhang} 2000)]
{Cheng00}
{Cheng}, K.~S., {Ruderman}, M., \& {Zhang}, L. 2000, \apj, {\bf 537}, 964

\bibitem[{Cusumano} {et~al.}(2003)]
{Cusumano03}
{Cusumano}, G., {Hermsen}, W., {Kramer}, M., {Kuiper}, L., {L{\"o}hmer}, O.,
  {Massaro}, E., {Mineo}, T., {Nicastro}, L., \& {Stappers}, B.~W. 2003, \aap,
  {\bf 410}, L9

\bibitem[({Daugherty} \& {Harding} 1996)]
{Daugherty96}
{Daugherty}, J.~K., \& {Harding}, A.~K. 1996, \aaps, {\bf 120}, C107

\bibitem[{Harding} {et~al.}(2002)]
{Harding02}
{Harding}, A.~K., {Strickman}, M.~S., {Gwinn}, C., {Dodson}, R., {Moffet}, D.,
  \& {McCulloch}, P. 2002, \apj, {\bf 576}, 376

\bibitem[{Johnston} {et~al.}(2001)]
{Johnston01}
{Johnston}, S., {van Straten}, W., {Kramer}, M., \& {Bailes}, M. 2001, \apjl,
  {\bf 549}, L101

\bibitem[{Krishnamohan} \& {Downs}(1983)]
{Krishnamohan83}
{Krishnamohan}, S., \& {Downs}, G.~S. 1983, \apj, {\bf 265}, 372

\bibitem[Lommen {et~al.}(2006)]
{Lommen06vela}
Lommen, A., Donovan, J., Gwinn, C., Arzoumanian, Z., Harding, A., Strickman, M., Dodson, R., 
McCulloch, P., \& Moffett, D. 2006, ApJ, accepted

\bibitem[{Petrova}(2003)]
{Petrova03}
{Petrova}, S.~A. 2003, \mnras, {\bf 340}, 1229

\bibitem[{Vivekanand}(2001)]
{Vivekanand01}
{Vivekanand}, M. 2001, \mnras, {\bf 326}, L33

\end{thebibliography}
\end{document}